# Contribution of top barrier materials to high mobility in near-surface InAs quantum wells grown on GaSb(001)

Joon Sue Lee[1], Borzoyeh Shojaei[1,2], Mihir Pendharkar[3], Mayer Feldman[4], Kunal Mukherjee[2], Chris J. Palmstrøm[1,2,3]

[1]California NanoSystems Institute, University of California, Santa Barbara, CA 93106
[2]Materials Department, University of California, Santa Barbara, CA 93106
[3]Department of Electrical and computer Engineering, University of California, Santa Barbara, CA 93106
[4]Physics Department, University of California, Santa Barbara, CA 93106

ABSTRACT

Near-surface InAs two-dimensional electron gas (2DEG) systems have great potential for realizing networks of multiple Majorana zero modes towards a scalable topological quantum computer. Improving mobility in the near-surface 2DEGs is beneficial for stable topological superconducting states as well as for correlation of multiple Majorana zero modes in a complex network. Here, we investigate near-surface InAs 2DEGs (13 nm away from the surface) grown on GaSb(001) substrates, whose lattice constant is closely matched to InAs, by molecular beam epitaxy. The effect of 10-nm-thick top barrier to the mobility is studied by comparing $Al_{0.9}Ga_{0.1}Sb$ and $In_{0.75}Ga_{0.25}As$ as a top barrier on otherwise identical InAs quantum wells grown with identical bottom barrier and buffer layers. A 3-nm-thick capping layer on $Al_{0.9}Ga_{0.1}Sb$ top barrier also affects the 2DEG electronic transport properties by modifying scattering from 2D remote ionized impurities at the surface. The highest transport mobility of 650,000 $cm^2$/Vs with an electron density of $3.81 \times 10^{11}$ $cm^{-2}$ was observed in an InAs 2DEG with an $Al_{0.9}Ga_{0.1}Sb$ top barrier and an $In_{0.75}Ga_{0.25}As$ capping layer. Analysis of Shubnikov-de Haas oscillations in the high mobility sample suggests that long-range scattering, such as remote ionized impurity scattering, is the dominant scattering mechanism in the InAs 2DEGs grown on GaSb(001) substrates. In comparison to InAs quantum wells grown on lattice-mismatched InP, the ones grown on GaSb show smoother surface morphology and higher quantum mobility. However, $In_{0.75}Ga_{0.25}As$ top barrier in InAs quantum well grown on GaSb limits the transport mobility by charged dislocations formed in it, in addition to the major contribution to scattering from the alloy scattering.

Low-dimensional semiconductors with a strong spin-orbit coupling, such as InAs and InSb, are suitable for developing spintronic devices [1,2] and for exhibiting topological superconducting phases that host Majorana zero modes when interfaced with superconductors [3–5]. Signatures of Majorana zero modes have been demonstrated in the 1D InAs and InSb nanowire-based systems with superconductors [6,7]. In addition, devices fabricated by top-down lithography on near-surface InAs two-dimensional electron gas (2DEG) systems with an epitaxial superconductor [8] have recently been reported to show signatures of Majorana zero modes [9]. The superconductor/2DEG platforms have great potential for fabrication of large-scale nanostructure networks consisting of multiple Majorana zero modes needed to realize a scalable topological quantum computer [10].

In the 2DEG-based systems, the 2DEG is required to be placed near the surface in order to achieve proximity-induced superconductivity, and epitaxial growth of a superconductor provides atomically clean, electronically transparent interface between the superconductor and the 2DEG. As a trade-off for the coupling between the superconductor and the 2DEG, the near-surface 2DEGs have relatively low mobility, mostly due to the scattering of electrons from surface states and surface defects. In the recently reported Al/InAs 2DEG systems grown on InP substrates [8,9], the mobility is just comparable to or less than that of 1D nanowires. The existence of the topological superconducting state in superconductor/semiconductor systems can be influenced by disorders, spin-orbit coupling, Zeeman spin splitting, superconducting proximity effect, and chemical potential in semiconductor. The most crucial ingredients for stable topological superconducting state are a high mobility (less disorders) and a strong spin-orbit coupling [11]. In this work, we focus on the first ingredient, the high mobility, by carefully modifying near-surface InAs quantum well heterostructures.

The large lattice-mismatch (3.3 %) between the InAs epitaxial layer and InP substrate, which has been used for substrates in the recent reports [8,9], results in a high number of dislocations and disorder that degrades the electronic transport properties of the InAs 2DEGs. A GaSb substrate is more closely lattice-matched to InAs (lattice mismatch: 0.6 %), suggesting that higher quality InAs quantum well heterostructures should be expected. Indeed, fewer dislocations and higher transport mobility of the order of $10^5$ cm$^2$/Vs up to $2.4 \times 10^6$ cm$^2$/Vs, with electrostatic gating, has been demonstrated in deep InAs 2DEGs grown on GaSb substrates, where scattering from the surface is relatively insignificant [12–15]. While deep 2DEGs (more than 25 nm away from surface) have been investigated, more studies on near-surface InAs 2DEGs (~10 nm away from surface) grown on GaSb substrates are needed. Here we study the effect of the top barrier (AlGaSb vs InGaAs) and capping layer (InGaAs vs GaSb) materials to the carrier density and the mobility in near-surface InAs 2DEGs grown on GaSb substrates.

We prepared three different near-surface InAs quantum well heterostructures, denoted as samples A, B, and C, on semi-insulating (at low temperature) GaSb(001) substrates by molecular beam epitaxy (MBE). The following describes growth details of the InAs quantum well heterostructures. First, the oxide on the GaSb substrate was desorbed at 540°C under Sb-overpressure in an MBE chamber with a base pressure ~$5\times10^{-11}$ Torr. Then, the substrate was cooled down to 500°C for growth of antimonide-based buffer layers: 100 nm GaSb, 5 nm AlSb, 300 nm $Al_{0.9}Ga_{0.1}As_{0.1}Sb_{0.9}$, and AlSb (2.5 nm)/GaSb (2.5 nm) superlattice (repeated by 10 times). The $Al_{0.9}Ga_{0.1}As_{0.1}Sb_{0.9}$ buffer layer is electrically insulating and closely lattice-matched to GaSb substrate [13]. The AlSb/GaSb superlattice helps to release strain in the buffer layer and results in smoother surface morphology [13]. After the superlattice growth, a 60 nm $Al_{0.9}Ga_{0.1}Sb$ bottom barrier for the quantum well was grown at the same temperature as for the buffer layers, and the substrate temperature was lowered to 480°C for growing 7 nm InAs quantum well layer. Samples A, B, and C have identical buffer layers, bottom barrier, and InAs layer as described above, with identical growth conditions. For the top barrier, two different ternary materials of $Al_{0.9}Ga_{0.1}Sb$ and $In_{0.75}Ga_{0.25}As$ were investigated. For samples A and B, the substrate temperature was kept at 480°C, and a 10-nm-thick $Al_{0.9}Ga_{0.1}Sb$ top barrier was grown, followed by a capping layer growth (3 nm GaSb capping for sample A and 3 nm $In_{0.75}Ga_{0.25}As$ capping for sample B). For sample C, the substrate temperature was further cooled down to 460°C after InAs layer growth, and a 13-nm-thick $In_{0.75}Ga_{0.25}As$ top barrier was grown.

Al-containing antimonide layers are easily oxidized in air, so it is necessary to add a capping layer on the $Al_{0.9}Ga_{0.1}Sb$ top barrier. Although the lattice mismatch between $In_{0.75}Ga_{0.25}As$ and $Al_{0.9}Ga_{0.1}Sb$ is larger than that between GaSb and $Al_{0.9}Ga_{0.1}Sb$, the capping layer thickness of 3 nm is thinner than the critical thickness of both GaSb and $In_{0.75}Ga_{0.25}As$. One of the advantages of having $In_{0.75}Ga_{0.25}As$ as a capping layer is that $In_{0.75}Ga_{0.25}As$ is robust against typical chemicals used in fabrication processes. Current fabrication recipes used in the recent reports of induced superconductivity and Majorana zero modes studies [9,16] may be useable on $In_{0.75}Ga_{0.25}As$ capped InAs 2DEG systems with antimonide barriers and buffers.

In the later part of this paper, a comparison is made between near-surface InAs 2DEGs grown on GaSb(001) substrates and on InP(001) substrates. On lattice-mismatched InP(001) substrate, an InAs quantum well (sample D) was grown on a step-graded buffer of metamorphic $In_xAl_{1-x}As$ layers, gradually adapting the lattice parameter of InP towards InAs (x = 0.52 to 0.81, 50 nm for each layer and 800 nm in total). The graded buffer needs to be grown at a low substrate temperature of 330˚C in order to regulate the formation of defects and to minimize surface roughness. This is followed by a 100-nm-thick $In_{0.81}Al_{0.19}As$ layer as the substrate temperature is ramped up to 480˚C. A delta-doping of Si with $2 \times 10^{12}$ cm$^{-2}$ sheet density was placed with a subsequent $In_{0.81}Al_{0.19}As$ spacer (6 nm) before the active region growth of the quantum well: $In_{0.75}Ga_{0.25}As$ (4 nm) bottom barrier, InAs layer (7 nm), and $In_{0.75}Ga_{0.25}As$ (10 nm) top barrier grown at 460˚C [17]. An epitaxial Al layer was grown after an overnight cool down with the substrate power turned off and all the cells in idle (lowest) temperatures. The Al layer was selectively wet-etched using Transene D etchant before electronic measurements. The InAs quantum well heterostructure stacks of the four samples investigated in this study as well as corresponding conduction band edges of the active region with the carrier density distribution based on the self-consistent Schrödinger-Poisson calculations, are shown in Fig. 1.

Longitudinal resistance $R_{xx}$ and Hall resistance $R_{xy}$ were measured at cryogenic temperatures using van der Pauw geometry with a perpendicular magnetic field up to 14 T. Magnetoresistance exhibits Shubnikov–de Haas (SdH) oscillations in longitudinal resistance. At higher magnetic fields, integer quantum Hall states are clearly seen with vanishing longitudinal resistance between the peaks in samples A and B (Fig. 2). Plateaus of transverse Hall resistance $R_{xy}$ are accompanied with the vanishing $R_{xx}$ where $1/R_{xy} = n \, (e^2/h)$. Here, $h$, $n$, and $e$ are the Plank constant, the integer filling factor ($n = 1, 2, 3…$), and the electron charge, respectively. Well-developed $R_{xy}$ plateaus are seen up to $n = 1$ within the measurement range in sample B with $In_{0.75}Ga_{0.25}As$ capping layer on $Al_{0.9}Ga_{0.1}Sb$ top barrier. The resulting sheet electron density $n_{Hall}$ is $3.81 \times 10^{11}$ cm$^{-2}$ and the transport mobility $\mu$ is 650,000 cm$^2$/Vs, at 0.5 K. The electron density is also confirmed by the SdH oscillations: $n_{SdH} = 3.87 \times 10^{11}$ cm$^{-2}$. The different capping layer of GaSb (sample A) results in a few times higher electron density $n_{Hall} = 1.33 \times 10^{12}$ cm$^{-2}$ with transport mobility $\mu = 287,000$ cm$^2$/Vs. The higher electron density may populate the first and the second electron subbands of the InAs 2DEG. At sufficiently high electron density, magneto-intermixing scattering between the first subband SdH series and other subband SdH series results in beat patterns in the magnetoresistance oscillations [18–20]. We observe the beating effect in sample A, as shown in Fig. 3(a). Fast Fourier transform (FFT) analysis of the SdH oscillations reveals the two peaks of closely separated oscillation frequencies [Fig. 3(a), inset]. In the lower density case (sample B), no beat pattern is seen [Fig. 3(b)], indicating that only the first electron subband is occupied in the InAs 2DEG with $In_{0.75}Ga_{0.25}As$ capping layer on $Al_{0.75}Ga_{0.25}As$ top barrier. The splitting of the magnetoresistance peaks appears due to Zeeman spin splitting above 1 T.

We now focus on analyzing the SdH oscillations from the first electron subband in sample B. Temperature dependence of amplitude of the SdH oscillations, given by subtracting a polynomial background from magnetoresistance $R_{xx}$, decreases as temperature increases [Fig. 3(c)]. The amplitude $\Delta R_{xx}$ of the SdH oscillations in small magnetic field is given by [21,22]

$$\Delta R_{xx} = 4R_0 X(T)\exp(-\pi/\omega_c \tau_q), \tag{1}$$

where $R_0$, $X(T)$, $\omega_c$, and $\tau_q$ are the zero-field resistance, a thermal damping factor, the cyclotron frequency, and the quantum lifetime, respectively. Here, $X(T) = (2\pi^2 k_B T/\hbar\omega_c)/\sinh(2\pi^2 k_B T/\hbar\omega_c)$ with $k_B$ being the Boltzmann constant, T being the sample temperature, and $\hbar\omega_c$ being the cyclotron energy. $\omega_c = eB/m^*$ with $e$ being the elementary electric charge, $B$ being the magnetic field, and $m^*$ being the effective mass. Analysis of the amplitude as a function of temperature determines the effective mass $m^*$. We obtained zero-field effective mass $m^* = (0.025 \pm 0.001)m_0$, where $m_0$ is the free electron mass. The result is broadly consistent with the previous experimental value from deep InAs/AlSb 2DEGs [12] as well as with a theoretical estimate using screened hybrid functional and spin-orbit coupling [23]. The quantum lifetime $\tau_q$ can be determined from the slope of a Dingle plot of $Ln(\Delta R_{xx}/4R_0 X(T))$ as a function of $1/B$, as shown in Fig. 3(d). The resulting quantum lifetime is $\tau_q = 0.21$ ps. In contrast, the Drude transport lifetime $\tau_t = \mu m^*/e$, inferred from the Hall mobility, is determined to be 9.24 ps. The ratio of the transport and quantum lifetime is rather large ($\tau_t/\tau_q = 45$), indicating the electronic transport is dominated by long-range scattering presumably from the remote ionized impurities [24,25].

Table I compares quantum well heterostructures and electronic transport characteristics of samples A through D. The electron density and the transport mobility shown in Table I were obtained from Hall measurements at 0.5 K for samples A, B, and C, using van der Pauw geometry, and 2 K for sample D, using Hall bar geometry. In the temperature range of 0.5–2 K, carrier density and the transport mobility remains almost identical (difference within 1% in sample B). The channel of the Hall bar (100-nm-wide and 400-nm-long) was defined by standard photolithography and wet-etching ($H_2O$ : citric acid : $H_3PO_4$ : $H_2O_2$ = 220 : 55 : 3 : 3). For calculation of density and mobility, zero-field longitudinal resistance $R_{xx}(B = 0\ T)$ and Hall resistance $R_{xy}(B)$ in a field range between -0.5 T and 0.5 T were used. Figure 4 plots the transport mobility of the four samples studied in this work as a function of the electron density. In addition, it includes three more data points (one black square and two red circles) from 2DEGs grown on InP(001) substrates: the black square is from the recent report [9], which studied signatures of Majorana zero modes; and the two red circles are from identical quantum well heterostructures to sample D but grown with some unintentional changes in the growth condition. We note that the red circles are from measurements using hall bar geometry, and the black square is from measurements using top-gated Hall bar geometry with zero gate voltage applied.

Differences in the mobility and the carrier density in samples A and B imply that a capping layer in the InAs quantum well heterostructures plays roles in addition to protection of the Al-containing top barrier. Both GaSb and In$_{0.75}$Ga$_{0.25}$As have a similar band gap of ~0.7 eV. However, difference in the Fermi energy pinning at the two surfaces and in the band alignment with Al$_{0.9}$Ga$_{0.1}$Sb top barrier, significantly modifies the conduction band profile $E_c$ of samples A and B [Figs. 1(e) and 1(f)]. The different conduction band profiles lead to different carrier densities based on the self-consistent calculations of the carrier density distribution of the lowest-energy electron wave function. With an assumption of identical remote ionized impurities in both capping layers, the carrier density of sample A is a few times higher than that of sample B, which is qualitatively consistent with the results from the transport measurements. Identical heterostructure stacks below the capping layer in samples A and B lead to the same contributions to scattering from alloy disorder and homogeneous background impurities, 3D remote ionized impurities located in the top and bottom barriers, charged dislocations, and interface roughness. The remaining contributions to scattering are from the surface of the capping layer, especially 2D remote ionized impurities located at the surface. Growth of 3-nm-thick pseudomorphic capping layer does not change the surface roughness in samples A and B, confirmed by atomic force microscopy (AFM) in Fig. 5. From the self-consistent calculation of the energy band structure and the carrier density distribution, the 2D remote ionized impurities at the surface in samples A and B are estimated to be ~5 × 10$^{12}$ cm$^{-2}$ and ~1 × 10$^{12}$ cm$^{-2}$, respectively.

The transport mobility in sample C (25,500 cm$^2$/Vs) is lower by an order of magnitude than that in sample B (650,000 cm$^2$/Vs). The main difference between samples B and C is the top barrier: 10-nm-thick Al$_{0.9}$Ga$_{0.1}$Sb with 3-nm-thick In$_{0.75}$Ga$_{0.25}$As capping layer in sample B and 13-nm-thick In$_{0.75}$Ga$_{0.25}$As with no additional capping layer in sample C. The band gap of Al$_{0.9}$Ga$_{0.1}$Sb (1.6 eV) is twice as large as that of In$_{0.75}$Ga$_{0.25}$As (0.7 eV). The electron wave function in sample B is highly confined in both upper and lower parts of the 2DEG while the electron wave function in the upper part of the 2DEG in sample C with the In$_{0.75}$Ga$_{0.25}$As top barrier is less confined [Figs. 1(f) and 1(g)]. Due to a larger portion of the electron wave function being present in the In$_{0.75}$Ga$_{0.25}$As top barrier in sample C, more alloy scattering in the top barrier is expected in sample C in comparison to sample B. Since the top most surface (In$_{0.75}$Ga$_{0.25}$As) is identical, the 2D remote ionized impurities at the surfaces of samples B and C should be comparable. The lattice constant of the In$_{0.75}$Ga$_{0.25}$As is smaller than those of the buffer layers that are closely matched to GaSb substrate, and the lattice mismatch of In$_{0.75}$Ga$_{0.25}$As to the buffer is ~2% with a critical thickness of ~15 nm. The thickness of the In$_{0.75}$Ga$_{0.25}$As top barrier (13 nm) is close to the critical thickness, so that misfit dislocations may start forming in the In$_{0.75}$Ga$_{0.25}$As layer in sample C. In contrast, the In$_{0.75}$Ga$_{0.25}$As capping layer (3 nm) in sample B is much thinner than the critical thickness, so that the contribution to scattering from the charged dislocations in the capping layer is expected to be negligible.

We now compare the InAs 2DEGs grown on GaSb substrates to an InAs 2DEG grown on InP substrate (sample D), which has similar heterostructure stacks to the ones used in the recent report of Majorana zero mode studies [9]. The surface morphology characterized by AFM shows that samples grown on GaSb substrates have smoother surface [Figs. 5(a) through 5(c)] while Sample D grown on an InP substrate reveals cross-hatch patterns that originate in the misfit dislocations that formed in the In$_{0.81}$Al$_{0.19}$As graded buffer [Fig. 5(f)]. The peak-to-valley height difference on the surface of sample D is an order of magnitude higher than those of the samples grown on GaSb substrates. We note that in sample D the surface morphology remains the same before and after Al etch. Rough surface morphology of sample D suggests additional contributions to scattering from the top most surface and interface roughness between InAs layer and the In$_{0.75}$Ga$_{0.25}$As top and bottom barriers.

Samples C and D were further characterized by plan-view electron channeling contrast imaging (ECCI) in a scanning electron microscope, which non-destructively reveals crystallographic defects over large areas. A three beam (g=02-2, 004) imaging condition was used for these samples. Lines with enhanced contrast correspond to misfit dislocations (light or dark depending on the Burgers vector of the dislocation), which are clearly seen in both samples (Fig. 6, electron channeling patterns in inset). The incident electron wavefront scatters and dampens as a function of depth into the crystal, resulting in a broadening and reduction of the defect-related backscattered electron contrast as the defect depth increases. Sharp lines were observed in sample C while much broader lines were observed in sample D, indicating that the misfit dislocations are located near the surface (active quantum well region) in sample C and deep in the graded In$_x$Al$_{1-x}$As buffer layers in sample D. From further analysis of the sharp lines in sample C, we found that there are two different types of lines: first type of lines has narrower full width at half maximum (FWHM) of ~100 nm with higher contrast [red curve in Fig. 6(a) inset], and the second type has relatively wider FWHM of ~200 nm with lower contrast [black curve in Fig. 6(a) inset]. The dislocation density of the first and second types is relatively low at $6 \times 10^3$ cm$^{-1}$ and $3 \times 10^3$ cm$^{-1}$, respectively. We attribute the first type of narrow lines to the misfit dislocations formed in the

In$_{0.75}$Ga$_{0.25}$As top barrier and the second type to the misfit dislocations formed in the Al$_{0.9}$Ga$_{0.1}$Sb bottom barrier. We note that the estimated misfit dislocation density of fully relaxed In$_{0.75}$Ga$_{0.25}$As and Al$_{0.9}$Ga$_{0.1}$Sb on GaSb is more than an order of magnitude greater at $3 \times 10^5$ cm$^{-1}$ and $1 \times 10^5$ cm$^{-1}$, respectively, suggesting that the active quantum well region was grown almost pseudomorphically and the misfit dislocations are an infrequent occurrence.

In sample D, the broad lines from the deep misfit dislocations overlap with each other, so the estimation of the misfit dislocation density in sample D is impractical. However, the lack of observation of sharp lines from ECCI of sample D confirms that the In$_{0.75}$Ga$_{0.25}$As top barrier in sample D is closely lattice-matched to the top most buffer layer of In$_{0.81}$Al$_{0.19}$As with 0.4 % lattice mismatch and ~75 nm critical thickness. Thus, dislocations lying far away from the InAs 2DEG and the interface roughness arising from these dislocations may not be the main contributions to scattering since mobility of around $1 \times 10^6$ cm$^2$/Vs has been demonstrated in a deep (120 nm away from surface) InAs quantum well grown on lattice-mismatched InP substrate with similar graded In$_x$Al$_{1-x}$As buffer layers [26]. Similar to the case of sample C, we attribute to remote ionized impurities and alloy scattering in the In$_{0.75}$Ga$_{0.25}$As top barrier the main scattering mechanisms in sample D. Since the electron wave function also resides in the In$_{0.75}$Ga$_{0.25}$As bottom barrier and upper part of In$_{0.81}$Al$_{0.19}$As buffer in sample D [Fig. 1(h)], alloy scattering is expected to be higher in sample D. However, the transport mobility in sample D is higher than that in sample C. It could be that contributions to scattering from the misfit dislocations in the top and bottom barriers of sample C may not be negligible after all.

One important point to note is that the onset magnetic field of SdH oscillations is much lower in Sample C (0.35 T) than in sample D (2.0 T). This implies that sample C has higher quantum mobility, which is inversely proportional to the onset magnetic field of the SdH oscillations, as well as longer quantum lifetime. The quantum mobility is sensitive to both small angle and large angle scattering while the transport mobility is sensitive only to large angle scattering. Our results suggest that small angle scattering is more pronounced in the InAs 2DEG grown on InP substrate.

In conclusion, we studied near-surface (10-13 nm away from surface) InAs 2DEGs grown on GaSb(001) substrates, which have a great promise to be employed in the superconductor/2DEG platforms for topological quantum computing applications. InAs 2DEGs with an Al$_{0.9}$Ga$_{0.1}$Sb top barrier of a wider band gap (1.6 eV) is shown to improve transport mobility by an order of magnitude compared to those with a narrower band gap In$_{0.75}$Ga$_{0.25}$As top barrier (0.7 eV). Capping layers (GaSb and In$_{0.75}$Ga$_{0.25}$As) on top of Al$_{0.9}$Ga$_{0.1}$Sb top barrier also affects the electronic properties. With 3-nm-thick In$_{0.75}$Ga$_{0.25}$As capping layer on InAs 2DEG with $Al_{0.9}Ga_{0.1}Sb$ top barrier, highest transport mobility of 650,000 cm$^2$/Vs and lowest electron density of $3.81 \times 10^{11}$ cm$^{-2}$ were observed. Based on the SdH analysis, effective mass $m^* = (0.025 \pm 0.001)m_0$ was obtained, and large transport lifetime to quantum lifetime ratio ($\tau_t/\tau_q = 45$) suggests long-range scattering is dominant in the InAs 2DEG with $Al_{0.9}Ga_{0.1}Sb$ top barrier grown on GaSb substrate. In comparison to InAs 2DEGs grown on InP substrates, InAs 2DEGs grown on GaSb substrates show smoother surface without cross-hatch patterns that are seen in samples grown on InP substrates. Although comparable transport mobility is observed in both InAs 2DEGs grown on GaSb and InP substrates with $In_{0.75}Ga_{0.25}As$ top barriers, onset magnetic field of SdH oscillations in the 2DEGs grown on GaSb substrate is lower by an order of magnitude, implying higher quantum mobility and longer quantum lifetime in InAs 2DEGs grown on GaSb substrates.

ACKNOWLEDGEMENTS

This work was supported by Microsoft Research Station Q. Part of this work was carried out in Nano-fabrication facility and the NSF funded Materials Research Science and Engineering Center at UCSB under Award No. DMR-1720256. We acknowledge Anthony P. McFadden for his support on electronic transport measurements.

TABLE I. Transport characteristics of the InAs 2DEGs.

| Sample | Substrate | Top barrier | Capping layer | Quantum well | $n_{2D,Hall}{}^a$ ($10^{11}$ cm$^{-2}$) | $\mu^a$ ($10^3$ cm$^2$/Vs) | Onset of SdH oscillations$^a$ (T) |
|---|---|---|---|---|---|---|---|
| A | GaSb(001) | AlGaSb 10 nm | GaSb 3 nm | InAs 7 nm | 13.3 | 287 | 0.38 |
| B | GaSb(001) | AlGaSb 10 nm | InGaAs 3 nm | InAs 7 nm | 3.81 | 650 | 0.2 |
| C | GaSb(001) | InGaAs 13 nm | - | InAs 7 nm | 8.12 | 25.5 | 0.35 |
| D | InP(001) | InGaAs 10 nm | - | InAs 7 nm | 8.70 | 52.4 | 2.0 |

$^a$Measured at 0.5 K using van der Pauw method except for sample D, which was measured at 2 K.

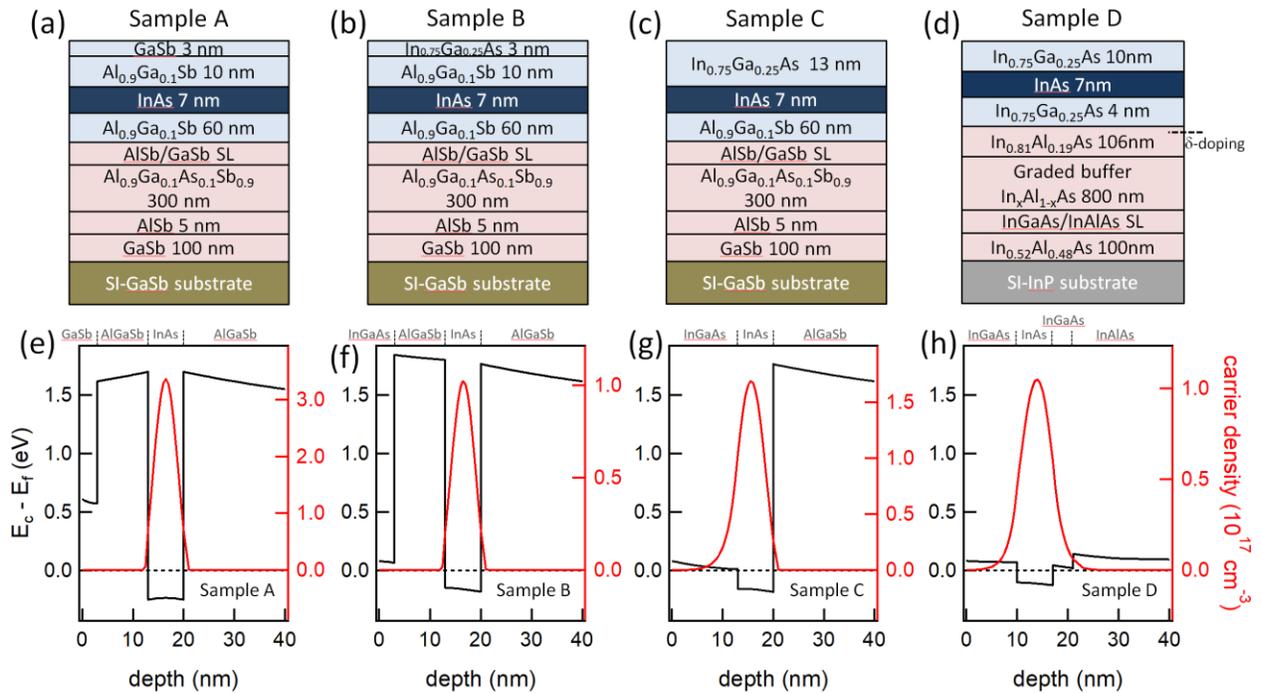

FIG 1. Schematics of the InAs 2DEG heterostructure stacks (a-c) grown on semi-insulating GaSb(001) substrates (Samples A, B, and C) and (d) grown on a semi-insulating InP(001) substrate (sample D). (e-h) Self-consistent calculations of the conduction band profile $E_c$ (black) from the Fermi level $E_f$ and the carrier density distribution (red) of the lowest electron level from Samples A, B, C and D.

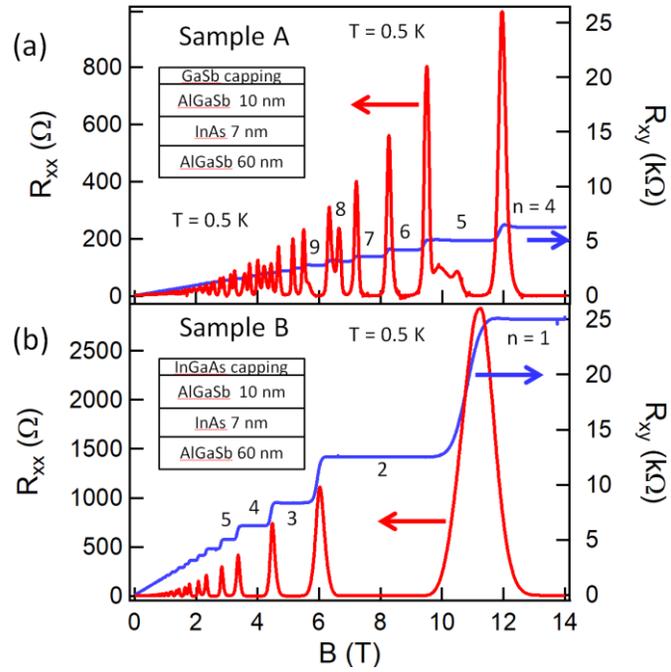

FIG 2. Longitudinal resistance $R_{xx}$ (red) and Hall resistance $R_{xy}$ (blue) from (a) sample A and (b) sample B show well-developed integer quantum Hall effect. All measurements were carried out at 0.5 K. Insets illustrate the active layers of the quantum wells.

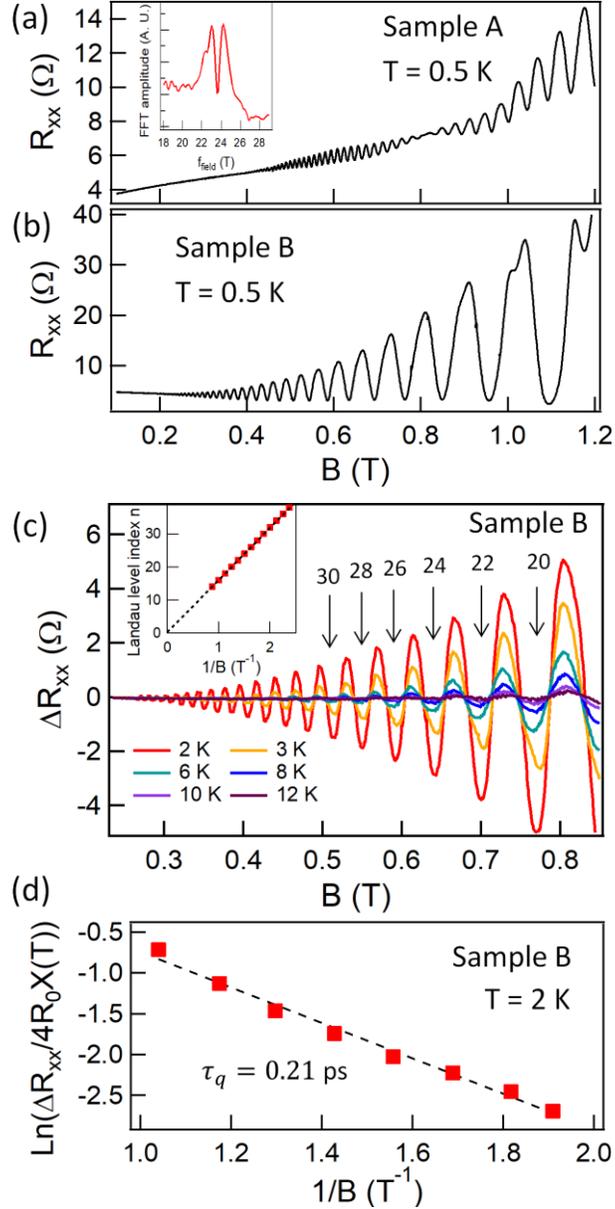

FIG 3. (a) Low-field magnetoresistance oscillations of sample A show a beat pattern. Inset: FFT analysis of the SdH oscillations reveals two peaks of closely separated oscillation frequencies. (b) Magnetoresistance oscillations of sample B. Above 1 T, the peak splits due to the Zeeman spin splitting. (c) Magnetoresistance oscillations of sample B after subtracting a polynomial background at various temperatures up to 12 K. Arrows indicate selected Landau level indices. Inset: Landau level index n as a function of 1/B. Black dashed line is a linear fit to data. (d) Dingle anaylsis of $Ln(\Delta R_{xx}/4R_0 X(T))$ as a function of $1/B$ from resistance oscillations at 2 K in (c). Black dashed line is a linear fit to data.

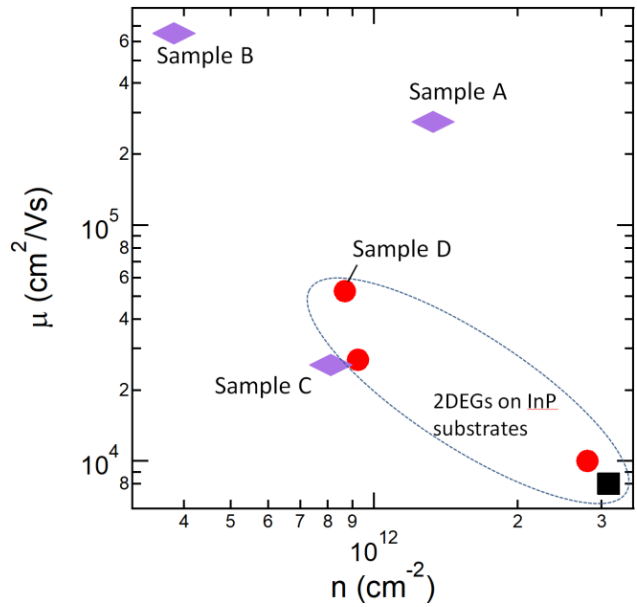

FIG 4. Transport mobility as a function of electron density from the four samples studied in the main text and additional InAs/In$_{0.75}$Ga$_{0.25}$As 2DEGs grown on InP substrates (red circles and a black square).

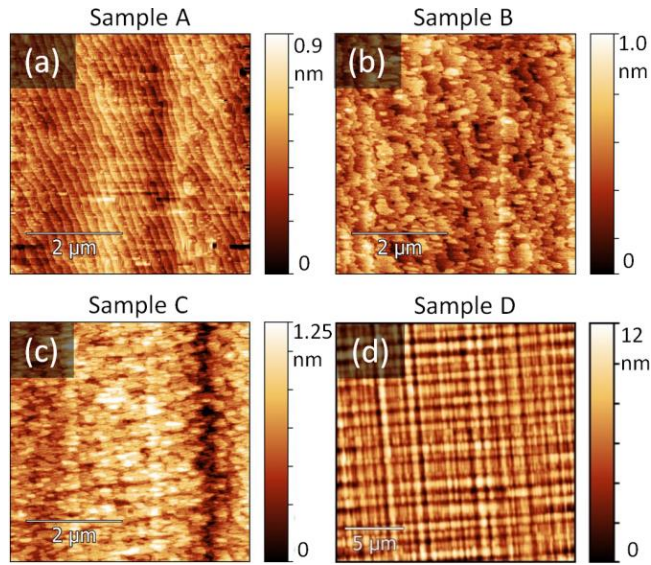

FIG 5. Schematics of InAs/In$_{0.75}$Ga$_{0.25}$As 2DEG heterostructure stacks on (a) a GaSb substrate (Sample C) and on (b) an InP substrate (sample D). 5×5 µm$^2$ AFM images of (c) sample A, (d) sample B, and (e) sample C grown on GaSb substrates reveal smooth surface morphology. (d) 20×20 µm$^2$ AFM images of sample D grown on InP substrate shows cross-hatch patterns.

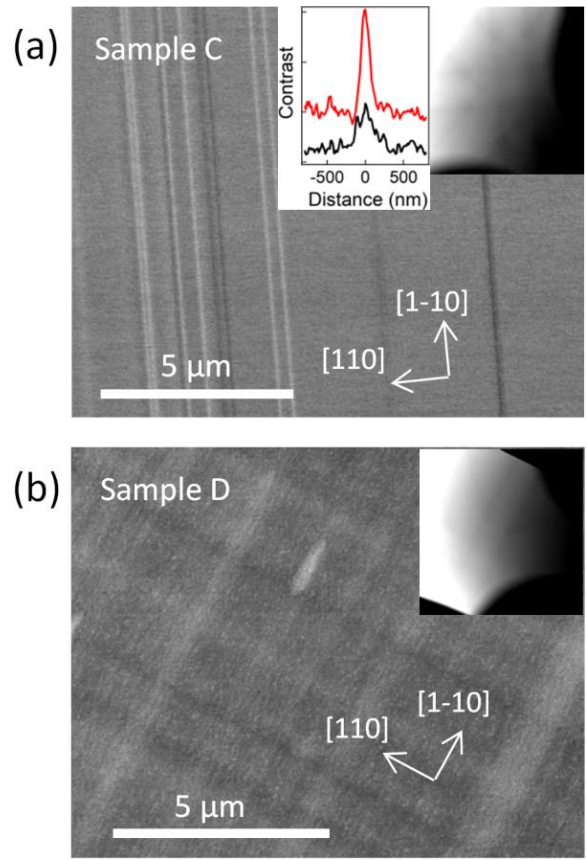

FIG 6. Electron channeling contrast images reveal the misfit dislocations in (a) sample C and (b) sample D. Inset plot in (a) shows two contrast profiles of representative narrower line (red) and wider line (black), corresponding to misfit dislocations in the top barrier and the bottom barrier in sample C, respectively. The electron channeling patterns are shown in insets in (a) and (b).